%% file: main.tex
\documentclass[letterpaper, 10 pt, conference]{ieeeconf} 
\IEEEoverridecommandlockouts 
\usepackage{amsmath,amsfonts}
\usepackage{nccmath}  
\usepackage{amssymb}
\usepackage{amsthm}
\usepackage{array}
\usepackage{graphicx}
\usepackage{textcomp}
\usepackage[justification=justified, font=small]{caption}
\usepackage{cuted}
\usepackage{standalone}
\usepackage{stfloats}
\usepackage{color}
\usepackage{url}
\usepackage{verbatim}
\usepackage[ruled,vlined]{algorithm2e}
\SetKw{Return}{Return}
\usepackage{tikz}
\usepackage{xcolor}
\usepackage{hyperref}
\hypersetup{colorlinks=true, urlcolor=blue, linkcolor=black, citecolor=black}
\usepackage{balance}
\usepackage{cite}
\usepackage{tikz}
\usetikzlibrary{arrows.meta,positioning,fit,calc,shapes.geometric,decorations.pathreplacing,backgrounds}
\usepackage{flushend}
\usepackage{courier}
\usepackage{listings}
\usepackage{booktabs}
\usepackage{multirow}
\usepackage[most]{tcolorbox}
\usepackage{url}
\usepackage[normalem]{ulem}
\usepackage{pgfplots}
\pgfplotsset{compat=1.18}

\definecolor{caucol}{RGB}{0,120,120}
\definecolor{aggcol}{RGB}{210,90,0}
\definecolor{gridgray}{RGB}{218,218,214}
\definecolor{nodecol}{RGB}{176, 226, 212}
\definecolor{egocol}{RGB}{31, 90, 190}
\definecolor{cInk}{HTML}{1A1A19}
\definecolor{cMuted}{HTML}{6B6B68}
\definecolor{cGrid}{HTML}{E3E3DF}
\definecolor{cParity}{HTML}{9A9A96}

\begin{document}

\title{GNN-Accelerated Mixed-Integer Dual MPC for Interactive Driving}
\author{Yidan Zhu$^{1,*}$, Shuhao Qi$^{2,*}$, Luyao Zhang$^{3}$, Sofie Haesaert$^{2}$, Jonas Mårtensson$^{1}$%
\thanks{$^{*}$ Indicates equally contribution.}%
\thanks{This work was supported by the Knut and Alice Wallenberg Foundation under the Wallenberg Artificial Intelligence, Autonomous Systems and Software Program (WASP), and by the European Union under the Horizon Europe project AIGGREGATE (Grant No.~101202457).}%
\thanks{$^{1}$ Yidan Zhu and Jonas Mårtensson are with the Division of Decision and Control Systems, School of Electrical Engineering and Computer Science (EECS), KTH Royal Institute of Technology, Stockholm, Sweden, and are also affiliated with the Integrated Transport Research Lab and Digital Futures.
 {\tt\small \{yidanz, jonas1\}@kth.se}}%
\thanks{$^{2}$ Shuhao Qi and Sofie Haesaert are with the Department of Electrical Engineering, Eindhoven University of Technology, Eindhoven, The Netherlands. {\tt\small \{s.qi, s.haesaert\}@tue.nl}}%
\thanks{$^{3}$ Luyao Zhang is with the Delft Center for Systems and Control, Delft University of Technology, Delft, The Netherlands. {\tt\small zly9706@gmail.com}}%
}

\maketitle

\begin{abstract} 



In interactions with uncertain opponents, dual model predictive control (MPC) can improve performance through information-seeking actions that reduce uncertainty about opponents' behavior. Its recent applications to autonomous driving, however, are limited to scenarios involving a single opponent on a single lane. 
This paper presents a mixed-integer dual MPC for multiple reactive opponents on multi-lane roads, jointly optimizing integer-valued maneuver decisions (lane changes and safe-region selections), and continuous motion over a scenario tree that samples plausible interactions with the opponents. 
As interaction complexity increases, solving the resulting mixed-integer nonlinear program becomes increasingly expensive. To reduce this computational burden, a graph neural network (GNN) predicts the optimal maneuver decisions, and high-confidence predictions are fixed before the reduced problem is solved. 
Simulations show that active probing behavior emerges in complex interactive scenarios, and that GNN guidance fixes $76.3\%$ of the integer decisions and reduces the solve time by $2.5\times$ on average, with negligible degradation of optimality.
\end{abstract}

\section{Introduction}
\begin{figure}[!t]
    \centering
    \includegraphics[width=0.9\columnwidth]{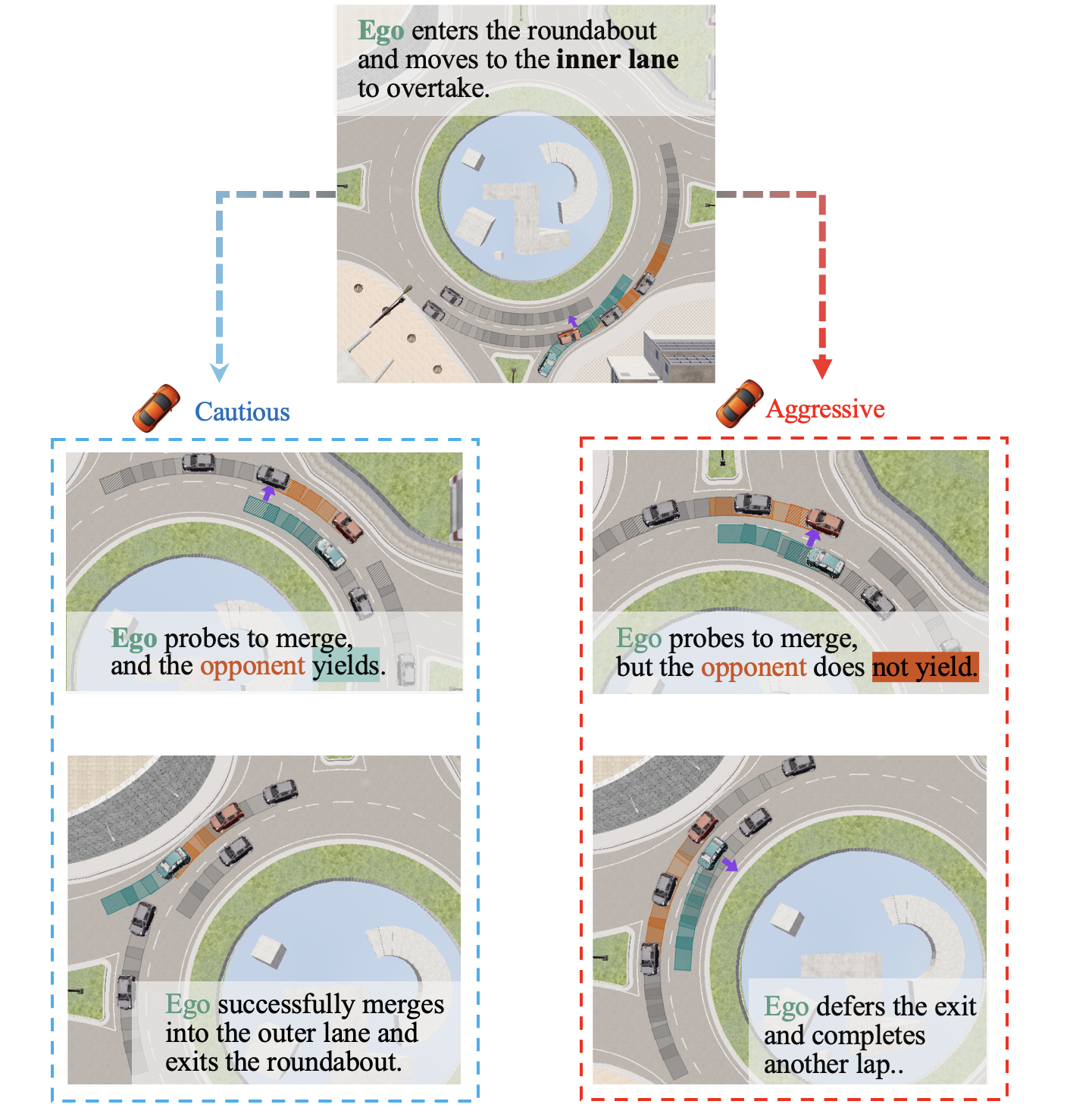}
    \caption{Interaction in a roundabout: the ego (green) enters the roundabout and moves to the inner lane to overtake, then probes to merge into the outer lane. If the opponent (orange) yields (cautious), the ego exits; otherwise (aggressive), it drives another lap.}
    \label{fig:roundabout} \vspace{-1mm}
\end{figure}

Advanced driver-assistance systems have been widely deployed in modern vehicles, yet fully autonomous driving still struggles in dense, interactive traffic~\cite{wang2022social}, exemplified by the roundabout in Fig.~\ref{fig:roundabout}. In such situations, the autonomous driving algorithm is often not confident enough to handle the risky interaction and requires the human driver to take over. What distinguishes interactive planning from conventional obstacle avoidance is that the surrounding vehicles are not only dynamic, but also reactive to the ego vehicle's motion~\cite{sadigh2018planning, qi2025situation}. The ego vehicle must therefore actively interact with other road users, rather than passively avoid them. For example, when merging into a dense traffic flow, as in the upper panel of Fig.~\ref{fig:roundabout}, a planner that does not account for the yielding reactions of other vehicles may never merge into the flow, an instance of the freezing robot problem~\cite{5654369}. Developing a controller that explicitly accounts for these reactions is thus critical for handling realistic traffic.

Model predictive control (MPC) is well suited to autonomous driving thanks to its native handling of constraints and uncertainty~\cite{mesbah2018stochastic}. Unlike \emph{Robust MPC}, which computes control inputs against the worst-case behavior of surrounding agents~\cite{zhou2022robust}, \emph{Branch MPC}~\cite{chen2021branch} mitigates the resulting conservatism by enumerating a scenario tree of plausible trajectories of the surrounding agents. However, these controllers can only adapt passively to the surrounding agents; they cannot proactively influence them. 
More recently, the \emph{dual-control effect}~\cite{knaup2024active, sadigh2018planning} has been introduced into MPC, termed \emph{Dual MPC}~\cite{hu2024active}, which couples estimation with control and generates behavior that simultaneously optimizes task performance and actively reduces uncertainty. In interactions with other agents, proactive behavior emerges from this coupling: the controller probes the opponent to reduce uncertainty about its hidden intent, and then exploits the resulting information to improve performance. Although Dual MPC exhibits such human-like proactive behavior, a gap remains in handling realistic interaction: due to the complexity of modeling multiple interacting agents and the structured road environment, existing approaches are often limited to a single surrounding vehicle and a single lane~\cite{hu2024active,qi2025situation}.

The planning problem in multi-lane, multi-vehicle traffic is combinatorial in nature, involving not only continuous trajectory optimization but also discrete maneuver decisions~\cite{oliveira2023interaction, 9329134}. For instance, the ego vehicle must decide, for each approaching opponent, whether to yield or pass and whether to bypass it on the left or on the right; on a structured road such as that in Fig.~\ref{fig:roundabout}, it must further choose whether to keep its lane, move to the inner lane to overtake, or move to the outer lane to exit. In practice, decision making and motion planning are commonly separated in a hierarchical framework~\cite{7490340}, but this decoupling loses information between the layers: the decisions are neither optimal nor timely, and the trajectories may fail to realize them in highly dynamic situations~\cite{9329134, quirynen2024real}. Thus, mixed-integer programming has been used to integrate the decision making and the trajectory optimization in a single optimization problem~\cite{9329134, quirynen2024real}, though without accounting for proactive interaction.

However, such a mixed-integer program is NP-hard in general~\cite{karp1972reducibility} and computationally intensive to solve~\cite{nemhauser1988integer}. The computational cost grows drastically with the complexity of the interactive scenario. In contrast, human drivers decide quickly even in complex interactive scenarios, thanks to experience and intuition. Learning-based methods bring a similar advantage to mixed-integer programs by exploiting the structure shared across instances to guide the solver. Graph neural networks (GNNs) have been used to learn branching policies~\cite{gasse2019exact}, to predict variable biases that guide the branch and bound (B\&B)~\cite{khalil2022mip}, and to generate partial integer assignments that define smaller subproblems~\cite{nair2020solving}.  

In this paper, we propose a dual MPC framework for interactive driving that accounts for the reactions of multiple surrounding vehicles and the structured road environment, and that generates reasonable interactive behavior, such as active information seeking, even in complex scenarios. To mitigate the computational burden, we design a learning-based pipeline that accelerates the solution of the resulting mixed-integer program with a GNN. Specifically, the GNN infers the optimal maneuver decisions to reduce the search space: those inferred with high confidence are fixed, and only the ambiguous ones are left to the B\&B search. The predicted decisions also make the complex interaction interpretable, which matters in this safety-critical application. With GNN guidance, the solve time is reduced by $2.5\times$ on average, with no loss of optimality. 


\section{Problem Formulation}
We present a problem formulation for interactive driving involving multiple vehicles and lanes. 

\subsection{System Model}
\label{subsec:system}
The ego vehicle, denoted by $V^e$, interacts with $N_o$ surrounding opponent
vehicles $\mathcal{V}^o\!=\!\{V^{o_m} \mid m\!=\!1,\dots,N_o\}$. Throughout, a superscript identifies the agent and a subscript the time step,
so that $(\mathbf{x}^e_t,\mathbf{u}^e_t)$ denotes the ego and
$(\mathbf{x}^{o_m}_t,\mathbf{u}^{o_m}_t)$ opponent $V^{o_m}$, with the label
shortened to $o$ whenever a single opponent is considered.
To keep the focus on the interaction among agents, every vehicle is modeled following~\cite{robbins2026mixed} as a simplified linear time-invariant system, a double integrator sampled at interval $\Delta t$,
\begin{equation}
\mathbf{x}^i_{t+1}=A\mathbf{x}^i_t+B\mathbf{u}^i_t,
\qquad \forall i\!\in\!\{e,o_1,\dots,o_{N_o}\},
\label{eq:dynamics}
\end{equation}
where the state
$\mathbf{x}^i\!=\![(\mathbf{p}^i)^\top\!,(\mathbf{v}^i)^\top]^\top\!\in\!\mathbb{R}^{4}$
collects the position $\mathbf{p}^i\!=\![p^i_x,p^i_y]^\top$ and the velocity
$\mathbf{v}^i\!=\![v^i_x,v^i_y]^\top$ in global Cartesian coordinates, the control
$\mathbf{u}^i\!=\![a^i_x,a^i_y]^\top\!\in\!\mathbb{R}^{2}$ is the
acceleration, and $A\!\in\!\mathbb{R}^{4\times4}$, $B\!\in\!\mathbb{R}^{4\times2}$
are the state-transition and input matrices, shared by all vehicles.

\subsection{Road Geometry and Path-Aligned Frames}
\label{subsec:road}

\begin{figure}[t]
    \centering
    \includegraphics[width=\linewidth]{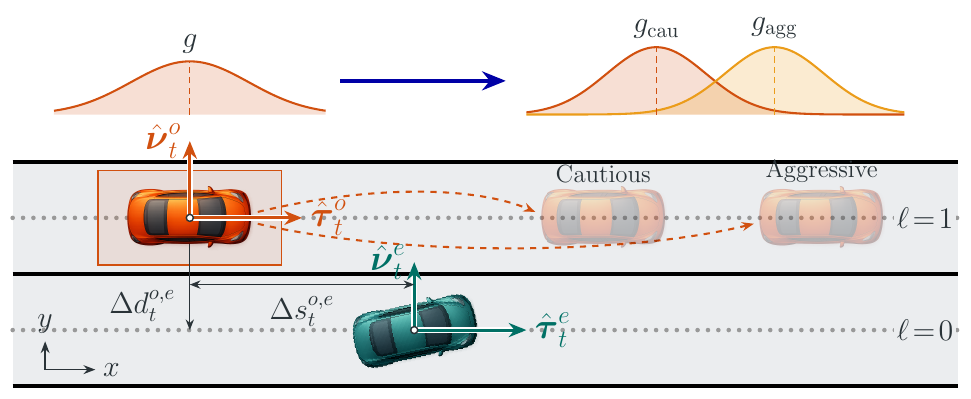}
    \caption{Interaction between the ego vehicle $V^e$ and an opponent $V^o$ on a multi-lane highway. The path-aligned frames $(\hat{\boldsymbol{\tau}}^i_t,\hat{\boldsymbol{\nu}}^i_t)$, $i\in\{e,o\}$, are anchored at the lane-centerline projections $\mathcal{T}^i_t$.} 
    \label{fig:highway_frames}
\end{figure}

Since driving unfolds along a road rather than in free space, we describe the motion of the vehicles relative to the lane structure of the road. The road has $N_\ell$ lanes of width $W$, indexed by $\ell\!\in\!\mathcal{L}\!:=\!\{0,\dots,N_\ell-1\}$. For each vehicle $i$ we build a path-aligned frame~\cite{werling2010optimal} on the centerline of the lane $\ell^{i}_t$ it currently occupies. Fig.~\ref{fig:highway_frames} illustrates the path-aligned frames for a highway interaction. To describe its geometry, that centerline is parameterized by arc length $s\!\in\![0,L]$, measured from the projection of the vehicle's position onto it, through a map $\mathcal{T}^{i}\colon[0,L]\!\to\!\mathbb{R}^2$ that takes $s$ to a planar position, where $L$ is the fixed length of the centerline segment. Its derivative is the unit tangent $\hat{\boldsymbol{\tau}}^{i}(s)\!=\!(\mathcal{T}^{i})'(s)$, and $\hat{\boldsymbol{\nu}}^{i}(s)$ the unit normal, its counterclockwise quarter turn. Anchored at $\mathcal{T}^{i}(s)$, this orthonormal pair defines the \emph{path-aligned} frame at $s$ and connects it to the global frame of~\eqref{eq:dynamics}. We write $(\hat{\boldsymbol{\tau}}^{i}_t,\hat{\boldsymbol{\nu}}^{i}_t)$ and $\mathcal{T}^{i}_t$ for the frame and anchor at the vehicle's own projection, $s\!=\!0$, at the current time $t$. 

The lateral offset of vehicle $i$ from the centerline of the lane it currently occupies is
\begin{equation}
\eta^{i}_t = (\hat{\boldsymbol{\nu}}^{i}_t)^{\top}\!\left( \mathbf{p}^{i}_t - \mathcal{T}^{i}_t \right).
\label{eq:lateral_proj}
\end{equation}
Accounting for its width $W^{i}$, vehicle $i$ stays within the road boundaries as long as $\underline{\eta}^{i}_t\!\le\!\eta^{i}_t\!\le\!\overline{\eta}^{i}_t$, where
\begin{equation}
\underline{\eta}^{i}_t\!:=\!\frac{W^{i}}{2}-\Big(\ell^{i}_t+\tfrac{1}{2}\Big)W, \;\;
\overline{\eta}^{i}_t\!:=\!\Big(N_\ell-\ell^{i}_t-\tfrac{1}{2}\Big)W-\frac{W^{i}}{2}.
\label{eq:road_bound}
\end{equation}
The same construction locates one vehicle relative to another: the longitudinal and lateral displacements and speeds of vehicle $j$ in the frame of vehicle $i$ are
\begin{equation}
\begin{aligned}
\Delta s^{i,j}_t&=(\hat{\boldsymbol{\tau}}^{i}_t)^\top\big(\mathbf{p}^{j}_t-\mathbf{p}^{i}_t\big), &\qquad
v^{i,j}_{\tau,t}&=(\hat{\boldsymbol{\tau}}^{i}_t)^\top\mathbf{v}^{j}_t,\\[2pt]
\Delta d^{i,j}_t&=(\hat{\boldsymbol{\nu}}^{i}_t)^\top\big(\mathbf{p}^{j}_t-\mathbf{p}^{i}_t\big), &\qquad
v^{i,j}_{\nu,t}&=(\hat{\boldsymbol{\nu}}^{i}_t)^\top\mathbf{v}^{j}_t.
\end{aligned}
\label{eq:frame_offsets}
\end{equation}
A vehicle's own longitudinal speed is abbreviated as $v^{i}_{\tau,t}\!:=\!v^{i,i}_{\tau,t}$. The ego vehicle must keep a safe distance from surrounding vehicles, which requires the longitudinal and lateral displacements of the ego relative to each opponent to satisfy:
\begin{equation}
\big|\Delta s^{o_m,e}_t\big|\ge d_\tau
\quad\text{or}\quad
\big|\Delta d^{o_m,e}_t\big|\ge d_\nu,
\label{eq:safe_problem}
\end{equation}
for all $m\!\in\!\{1,\dots,N_o\}$ and all $t\!\ge\!0$, where the longitudinal and lateral margins $d_\tau,d_\nu\!>\!0$ account for the vehicle length and width, respectively, plus a safety buffer. 

\subsection{Opponent Behavior Model}
\label{subsec:opponent}
To account for the reactions of the surrounding vehicles, we model how each opponent $V^{o_m}$ responds to the ego vehicle's motion. For simplicity, this paper ignores the interactions among opponents. Following the driver models in~\cite{hu2024active, 9294176, wang2019learning}, we adopt a parametric model that emulates realistic driving through three features: responsiveness to the ego vehicle, a latent intention parameter that is not directly observable, and stochasticity. Thus, the opponent's control input at time step $t$ is modeled as
\begin{equation}
\mathbf{u}^{o_m}_t=g(\mathbf{x}^{o_m}_t, \mathbf{x}^e_t;\theta^m)+\boldsymbol{\epsilon}^{o_m}_t,
\qquad \boldsymbol{\epsilon}^{o_m}_t\sim\mathcal{N}(\mathbf{0},\Sigma),
\label{eq:opp_general}
\end{equation}
where $g(\cdot)$ is a parametric policy depending on the opponent's own state and on the ego state $\mathbf{x}^e_t$, $\theta^m$ is the latent intention parameter of opponent $V^{o_m}$, and $\boldsymbol{\epsilon}^{o_m}_t$ is zero-mean Gaussian noise with covariance $\Sigma$, independent across vehicles and time steps, capturing behavioral variability.



\begin{tcolorbox}[breakable, enhanced jigsaw, colback=white,
 colframe=black, boxrule=0.4pt, arc=0pt,
 left=5pt, right=5pt, top=5pt, bottom=5pt, before skip=4pt, after skip=6pt]
\textbf{An instantiation of the opponent model}:
Here we present a simple model for an opponent $V^{o}$ with a scalar latent parameter $\theta$, interacting with the ego vehicle $V^e$. The policy $g(\cdot)$ of~\eqref{eq:opp_general} drives the opponent toward a commanded speed along its own path,
\begin{equation}
g\big(\mathbf{x}^{o}_t,\mathbf{x}^e_t;\theta\big)
=K_p\big(v^{\mathrm{cmd}}_t-v^{o}_{\tau,t}\big)\,
\hat{\boldsymbol{\tau}}^{o}_t,
\label{eq:opp_mean}
\end{equation}
with constant gain $K_p\!>\!0$. 
$V^{o}$ reacts to the ego only when the ego lies in its interaction region. Since we assume that each opponent keeps its lane, the \emph{interaction region} is the strip ahead of it: the ego lies in it if $0\!<\!\Delta s^{o,e}_t\!\le\!d_{\text{int}}$ and $|\Delta d^{o,e}_t|\!\le\!W_{\text{int}}$, where $d_{\text{int}}\!>\!0$ is its longitudinal range and $W_{\text{int}}\!>\!W$ its lateral half-width. The ego's speed and displacements in the opponent's frame follow from~\eqref{eq:frame_offsets} as $v^{o,e}_{\tau,t}$, $\Delta s^{o,e}_t$, and $\Delta d^{o,e}_t$. If the ego lies outside the interaction region, $V^{o}$ cruises at its desired speed, $v^{\mathrm{cmd}}_t\!=\!v^{o}_{\mathrm{des}}$; if the ego lies inside it, the commanded speed is
\begin{equation}
v^{\mathrm{cmd}}_t\!=\!\left\{\begin{array}{@{}l@{\;\;}l@{}}
v^{o,e}_{\tau,t}+\theta\,\Delta v,
 & |\Delta d^{o,e}_t|>W/2,\\
v^{o,e}_{\tau,t}\!+\!K_g(\Delta s^{o,e}_t\!-\!d_{\text{des}}),
 & |\Delta d^{o,e}_t|\le W/2,
\end{array}\right.
\label{eq:opponent_policy}
\end{equation}
where $d_{\text{des}}\!>\!0$ is the desired gap to the vehicle ahead, $K_g\!>\!0$ the gap gain, $v^{o}_{\mathrm{des}}$ the desired speed, $\Delta v\!>\!0$ the speed offset, and $\theta\!\in\!\Theta\!=\!\{\theta_{\mathrm{cau}},\theta_{\mathrm{agg}}\}\!=\!\{-1,+1\}$ the latent parameter. The first case describes the situation where the ego is outside the lane of $V^{o}$ and may merge into it, as shown in Fig.~\ref{fig:highway_frames}: if $V^{o}$ is \textit{cautious} ($\theta\!=\!\theta_{\mathrm{cau}}$), it reduces its speed below that of the ego and opens a gap; if $V^{o}$ is \textit{aggressive} ($\theta\!=\!\theta_{\mathrm{agg}}$), it targets a higher speed to close it. The second case applies when the ego occupies the lane ahead of $V^{o}$ and has become its leader: the opponent regulates its speed to maintain the gap $d_{\text{des}}$.
\end{tcolorbox}

\subsection{Problem Statement}
Given the reactive and uncertain opponents modeled by~\eqref{eq:opp_general}, the ego cannot simply predict their motion and plan around it. The hidden parameters $\theta^m$ that determine how each opponent responds are never observed directly. They are revealed only through the responses themselves, and those responses are triggered by the ego's own motion. A controller is therefore required that accounts for this coupling between estimation and control while tracking a preferred lane $\ell^{\mathrm{pref}}\!\in\!\mathcal{L}$ at a desired longitudinal speed $v^{e}_{\mathrm{des}}$ and satisfying the safety constraint~\eqref{eq:safe_problem} under the ego dynamics~\eqref{eq:dynamics}. It should operate in a setting with $N_o$ reactive opponents and $N_\ell$ lanes, rather than in the single-lane, single-opponent abstraction adopted in most prior work~\cite{qi2025situation,hu2024active}. Finally, the growth of the computational cost with the number of opponents and lanes should be mitigated for scalability.

\section{Mixed-Integer Dual MPC}
Unlike the prior work on dual MPC in~\cite{qi2025situation}, which is restricted to a single opponent in a single lane, we account for multiple lanes and multiple opponents by carrying the maneuver decisions as integer variables. These decisions are optimized jointly with the continuous motion over a scenario tree that enumerates the plausible future interactions. Along that tree, a belief over the latent intention parameter of~\eqref{eq:opp_general} is propagated and updated from the observed actions. The remainder of this section specifies this mixed-integer dual MPC formulation.
Following~\cite{xing2022vehicle,reiter2023frenet}, we retain both the Cartesian and the path-aligned frames and express each objective and constraint term in the more convenient one, with the vehicle states and their dynamics staying in the global Cartesian frame of~\eqref{eq:dynamics} while the reference-tracking terms and the lane-structure constraints are expressed in path-aligned coordinates.

\subsection{Integer Decision Variables}
The combinatorial nature of planning for autonomous driving arises from two sources: the road structure, which forces a choice of the lane to occupy, and the surrounding vehicles, which forces a choice of the region to occupy relative to each of them, namely passing, merging ahead, or following behind. We represent both choices as integer variables, so that the continuous motion and the maneuver decisions are optimized jointly. The lane choice is formulated first, followed by the choice of the convex safe region to occupy relative to each opponent.

\subsubsection{Lane-Change Decisions}\label{subsubsec:lane} At time $t$ the ego occupies lane $\ell^e_t\!\in\!\mathcal{L}$. To describe the lane-change decision, we define two binary variables $\delta^{+}_t, \delta^{-}_t\!\in\!\{0,1\}$, marking a transition to the adjacent lane of higher and of lower index, respectively, following the mixed-integer lane encoding of~\cite{quirynen2024real}. The difference $(\delta^{+}_t-\delta^{-}_t)\!\in\!\{-1,0,1\}$ is the signed lane change, negative for a move to the lower-index lane, zero for no change, and positive for a move to the higher-index lane. Since the ego cannot move to both adjacent lanes at once, the sum $\delta^{+}_t+\delta^{-}_t$ is bounded by one; the lane index then advances by this signed change,
\begin{equation}
\delta^{+}_t + \delta^{-}_t \le 1,
\quad
\ell^e_t = \ell^e_{t-1} + (\delta^{+}_t - \delta^{-}_t) \in\mathcal{L}.
\label{eq:lane_recursion}
\end{equation}

\subsubsection{Safe-Region Selections}
\label{sssec:relpos}
\begin{figure}[tb]
 \centering
\includegraphics[width=0.65\linewidth]{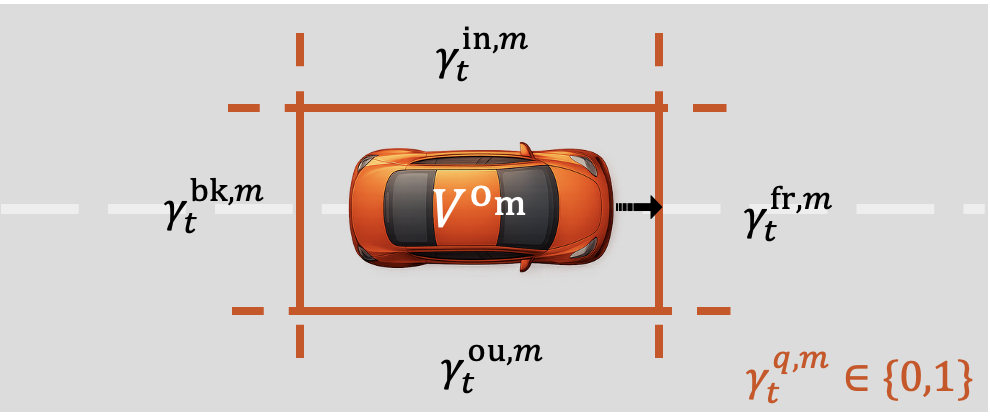}
 \caption{Illustration of the four convex safe regions (front, back, inner, outer) used for mixed-integer obstacle avoidance. } 
 \label{fig:safety_regions}
\end{figure}

For each opponent $V^{o_m}$, the collision-free set of relative positions admitted by~\eqref{eq:safe_problem} is non-convex. Following~\cite{richards2002aircraft}, we decompose it into four \emph{convex safe regions}, one per maneuver $q\!\in\!\mathcal{Q}\!=\!\{\mathrm{fr},\mathrm{bk},\mathrm{in},\mathrm{ou}\}$: passing ahead of the opponent, following behind it, and passing on its inner or outer side, the side of the positive or the negative normal $\hat{\boldsymbol{\nu}}^{o_m}_t$ (Fig.~\ref{fig:safety_regions}). Binary selectors $\gamma^{q,m}_t$, one per region, mark which region the ego occupies relative to $V^{o_m}$ at time $t$. By~\eqref{eq:frame_offsets}, the ego's position relative to that opponent is $\mathbf r^m_t\!=\!\big[\,\Delta s^{o_m,e}_t,\;\Delta d^{o_m,e}_t\,\big]^{\top}$, and each region is the halfspace $\mathbf a_q^{\top}\mathbf r^m_t\!\ge\!d_q$, where $\textstyle \mathbf a_{\mathrm{fr}}\!=\!-\mathbf a_{\mathrm{bk}}\!=\![1,0]^{\top}$ and $\mathbf a_{\mathrm{in}}\!=\!-\mathbf a_{\mathrm{ou}}\!=\![0,1]^{\top}$ select the longitudinal or the lateral offset, and $d_{\mathrm{fr}}\!=\!d_{\mathrm{bk}}\!=\!d_\tau$ and $d_{\mathrm{in}}\!=\!d_{\mathrm{ou}}\!=\!d_\nu$ are the margins of~\eqref{eq:safe_problem}. The selection is enforced via a big-$M$ method~\cite{richards2002aircraft},
\begin{equation}
\begin{aligned}
\textstyle &\sum_{q\in\mathcal{Q}} \gamma^{q,m}_t = 1, \\
& \mathbf a_q^{\top}\mathbf r^m_t \;\ge\; d_q - M\big(1-\gamma^{q,m}_t\big),
 \quad \forall q\!\in\!\mathcal{Q},
\end{aligned}
\label{eq:bigM}
\end{equation}
for every opponent $m$ and every time step $t$. The selector turns its halfspace on or off: at $\gamma^{q,m}_t\!=\!1$ the region is imposed as written, while at $\gamma^{q,m}_t\!=\!0$ a large enough $M$ relaxes the bound past anything the ego can reach, so that region places no restriction. With the one-hot constraint, exactly one halfspace is imposed, expressing the non-convex union through linear constraints.
With the above definitions, the ego's binary decisions at time $t$ are collected as
\begin{equation}
  \boldsymbol{\sigma}^e_t\!=\!\big(\delta^{+}_t,\delta^{-}_t,\{\gamma^{q,m}_t\}_{q\in\mathcal{Q},\,m=1,\dots,N_o}\big)\!\in\!\{0,1\}^{2+4N_o}.
  \label{eq:binary_decision}
\end{equation}

\subsection{Belief Representation and Update}
\label{subsec:belief}

Since the hidden parameter $\theta^m$ of opponent $V^{o_m}$ is not directly observable, the ego maintains a belief over $\theta\!\in\!\Theta$, $b^m_t(\theta)\!=\!\Pr\big(\theta^m\!=\!\theta\mid\{\mathbf{x}^e_{0{:}t},\mathbf{x}^{o_m}_{0{:}t}\}\big)$, where $\Pr(\cdot\mid\cdot)$ denotes the conditional probability given the ego and opponent states observed up to time $t$. The belief is normalized so that $\sum_{\theta\in\Theta}b^m_t(\theta)\!=\!1$. As the interaction proceeds, the ego observes the opponent's next state and recovers its realized input from the velocity increment, $\tilde{\mathbf{u}}^{o_m}_t\!=\!(\mathbf{v}^{o_m}_{t+1}\!-\!\mathbf{v}^{o_m}_t)/\Delta t$, from which the belief is updated by Bayes' rule,
\begin{equation}
b^m_{t+1}(\theta)=\frac{\rho_\theta(\tilde{\mathbf{u}}^{o_m}_t\mid\mathbf{x}^{o_m}_t,\mathbf{x}^e_t)\,b^m_{t}(\theta)}
{\sum_{\theta'\in\Theta}\rho_{\theta'}(\tilde{\mathbf{u}}^{o_m}_t\mid\mathbf{x}^{o_m}_t,\mathbf{x}^e_t)\,b^m_{t}(\theta')}.
\label{eq:online_belief}
\end{equation}
Here $\rho_\theta(\tilde{\mathbf{u}}^{o_m}_t\!\mid\!\mathbf{x}^{o_m}_t,\mathbf{x}^e_t)\!=\!\mathcal{N}\big(\tilde{\mathbf{u}}^{o_m}_t\!\mid\! g(\mathbf{x}^{o_m}_t,\mathbf{x}^e_t;\theta),\Sigma\big)$ is the likelihood of the observed input under the opponent policy~\eqref{eq:opp_general} with candidate value $\theta$, so each observation rescales the prior by how well it matches each candidate value. We write the update as $b^m_{t+1}\!=\!\Psi(b^m_{t},\tilde{\mathbf{u}}^{o_m}_t,\mathbf{x}^{o_m}_t,\mathbf{x}^e_t)$ with the operator $\Psi(\cdot)$.


\subsection{Scenario Tree}
\label{subsec:tree}

\begin{figure}[!t]
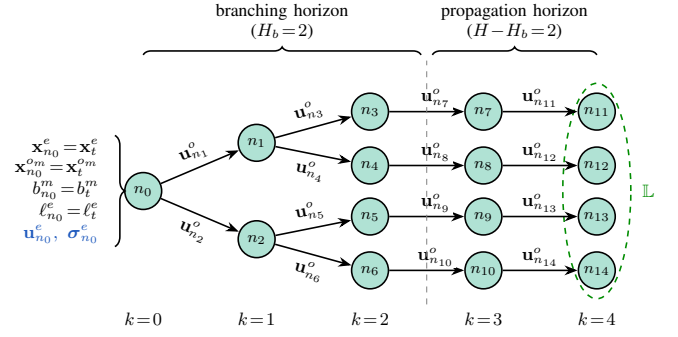

 \centering
 \includestandalone[width=\linewidth]{drawing/example}
 \caption{Example of a scenario tree with $H_b\!=\!2$ and $H\!=\!4$. The root $n_0$ is initialized from the current time $t$, and the edge into a node $n$ carries the sampled opponent reaction $\mathbf{u}^{o}_{n}$. The blue $\mathbf{u}^e_n$ and $\boldsymbol{\sigma}^e_n$ are the decision variables at that node.}
 \label{fig:tree}
\end{figure}

To capture the multi-modal uncertainty of the interaction, the controller plans over a scenario tree~\cite{chen2021branch}, which enumerates several plausible futures over a prediction horizon of $H$ steps ahead of the current time $t$, indexed by $k\!=\!0,\dots,H$. As shown in Fig.~\ref{fig:tree}, the tree has node set $\mathbb{N}\!=\!\{n_0,\dots,n_{N-1}\}$: the root $n_0$ is the current joint state at depth $k\!=\!0$, the leaves $\mathbb{L}\!\subset\!\mathbb{N}$ are the terminal nodes at depth $H$, and $\mathcal{P}(n)$, $\mathcal{C}(n)$ denote the parent and children of a node. The depth of a node is the step $k$ it predicts. Branching at every step would grow the node count exponentially with the horizon, so, to keep a long horizon tractable, the tree is generated in two phases~\cite{hu2024active}: a \emph{branching horizon} over the first $H_b$ steps, in which every node has several children, one per sampled joint reaction of the interacting opponents, followed by a \emph{propagation horizon} over the remaining steps up to $H$, in which every node has a single child. 

As shown in Fig.~\ref{fig:tree}, every node $n$ carries the joint state $(\mathbf{x}^e_n,\{\mathbf{x}^{o_m}_n\}_{m=1}^{N_o})$, the beliefs $\{b^m_n\}_{m=1}^{N_o}$, and the ego's lane index $\ell^e_n$, together with the decisions taken there, the binary decisions $\boldsymbol{\sigma}^e_n$ and the input $\mathbf{u}^e_n$. The edges carry the uncertainty about the opponents: the joint opponent input $\mathbf{u}^{o}_n\!=\!(\mathbf{u}^{o_1}_n,\dots,\mathbf{u}^{o_{N_o}}_n)$ on the edge into $n$ is one realization of~\eqref{eq:opp_general} under a sampled hidden parameter $\theta^{m}_n$ and a noise draw $\boldsymbol{\epsilon}^{o_m}_n$ for each opponent on that edge; both are drawn before the optimization and enter~\eqref{eq:full_problem} as data. The states $\mathbf{x}^e_n$ and $\mathbf{x}^{o_m}_n$ follow from the parent through the dynamics~\eqref{eq:dynamics}, the beliefs $b^m_n$ through the update~\eqref{eq:online_belief}, and the lane index $\ell^e_n$ through the recursion~\eqref{eq:lane_recursion}, with the consecutive time steps in these equations replaced by the parent $\mathcal{P}(n)$ and the node $n$. At node $n$, the input $\mathbf{u}^e_n$ and the binary decisions $\boldsymbol{\sigma}^e_n$ are optimized against the realizations at all its children $\mathcal{C}(n)$.

The quantities of Sec.~\ref{subsec:road} are expressed in path-aligned frames and used in the road bound~\eqref{eq:road_bound}, the safe-region selection~\eqref{eq:bigM}, the opponent behavior model~\eqref{eq:opp_mean}, and the ego's tracking objective defined below. Their frames are fixed before the solve: the frame of each vehicle $i$ lies on the centerline of the lane $\ell^{i}_t$ it occupies at the current time $t$ and the frame of horizon step $k$ is taken at the nominal arc length $s^{i}_k$ along it, $(\hat{\boldsymbol{\tau}}^{i}_k,\hat{\boldsymbol{\nu}}^{i}_k)\!=\!\big(\hat{\boldsymbol{\tau}}^{i}(s^{i}_k),\hat{\boldsymbol{\nu}}^{i}(s^{i}_k)\big)$ anchored at $\mathcal{T}^{i}_k\!=\!\mathcal{T}^{i}(s^{i}_k)$. Since every path-aligned quantity is a projection of the Cartesian states of~\eqref{eq:dynamics}, the frames of different vehicles are linked only through the global frame.

\subsection{Control Objective}
The control objective is a cost over the scenario tree, assembled from the terms introduced below.

\subsubsection{Lane Selection} The ego often has to keep to a particular lane to reach its destination, such as the exit lane of a roundabout. We therefore charge the assigned lane at each node for its distance from a preferred lane $\ell^{\mathrm{pref}}$, $J^{\mathrm{pref}}_n(\ell^e_n)\!=\!\lambda_{\mathrm{pref}}\big(\ell^e_n-\ell^{\mathrm{pref}}\big)^2$, with weight $\lambda_{\mathrm{pref}}\!\ge\!0$. When the ego leaves $\ell^{\mathrm{pref}}$ to overtake a vehicle, this term draws it back once the overtaking is complete. 


\subsubsection{Reference Tracking} The ego is expected to track the centerline of the assigned lane $\ell^e_n$ at a desired speed $v^{e}_{\mathrm{des}}$. Since the path-aligned frame of Sec.~\ref{subsec:road} stays on the centerline of the lane $\ell^e_t$ the ego currently occupies, the deviation from the assigned lane is obtained by shifting the lateral offset by the lane difference. At a node $n\!\in\!\mathbb{N}$ of depth $k$, the ego's lateral offset $\eta^e_n$ and speed along the road $v^e_{\tau,n}$ are given by~\eqref{eq:lateral_proj} and~\eqref{eq:frame_offsets} with the state at node $n$ and the frame of step $k$, and the tracking error for lane keeping and longitudinal speed there is
\begin{equation}
\mathbf{e}_n=\begin{bmatrix} \Delta\eta^{e}_n \\[2pt] v^e_{\tau,n}-v^{e}_{\mathrm{des}}\end{bmatrix},
\qquad \Delta\eta^{e}_n=\eta^e_n-\left(\ell^e_n-\ell^e_t\right)W ,
\label{eq:track_error}
\end{equation}
where $\Delta\eta^{e}_n$ is the deviation of the ego from the centerline of the assigned lane. Thus, the stage and terminal costs are the quadratic tracking terms, $J_n(\mathbf{e}_n,\mathbf{u}^e_n)\!=\!\|\mathbf{e}_n\|_Q^2\!+\!\|\mathbf{u}^e_n\|_R^2$ and $J^F_n(\mathbf{e}_n)\!=\!\|\mathbf{e}_n\|_{Q_f}^2$, with $Q,Q_f,R\!\succeq\!0$ in $\mathbb{R}^{2\times2}$.

\subsubsection{Safety Constraint Relaxation}
Since the scenario tree randomly samples several opponent reactions, some of them improbable, the ego may be unable to satisfy the safety constraints strictly for all of them. Thus, the road bound~\eqref{eq:road_bound} and the safe region selected in~\eqref{eq:bigM} are softened by nonnegative slacks: $\xi^{\mathrm{rd}}_n$ widens the road bound on both sides, and $\xi^{\mathrm{sf},m}_n$ lowers the bound $d_q$ of the region selected for opponent $m$. Stacked as $\boldsymbol{\xi}_n\!=\!\big[\xi^{\mathrm{rd}}_n,\xi^{\mathrm{sf},1}_n,\dots,\xi^{\mathrm{sf},N_o}_n\big]^{\top}\!\ge\!0$, the slacks are penalized at each node as $J^{\mathrm{slack}}_n(\boldsymbol{\xi}_n)=\boldsymbol{\lambda}_{\mathrm{slack}}^{\top}\boldsymbol{\xi}_n$, 
with $\boldsymbol{\lambda}_{\mathrm{slack}}\!>\!0$ chosen to strongly penalize constraint violations relative to tracking errors.

\subsubsection{Node Weighting} The behaviors sampled at the nodes are not equally probable, so the node weights $\omega_n$ are computed from the propagated belief on the tree. Since the noises in~\eqref{eq:opp_general} are independent across vehicles, the opponents' inputs along an edge are independent conditioned on the parent state, and the likelihood of the joint action $\mathbf{u}^{o}_n$ on that edge is the product of the per-opponent likelihoods. Along the tree, the belief over the hidden parameter of each opponent is propagated by the update~\eqref{eq:online_belief} as the interaction evolves with the sampled opponent reactions. Marginalizing each factor over the corresponding parent belief, and then normalizing over the children of the same parent, turns that likelihood into a parent-to-child transition probability, from which the node weight follows recursively from the root $\omega_{n_0}=1$,
\begin{equation}
\begin{aligned}
\bar\rho_n&=\prod_{m=1}^{N_o}\sum_{\theta\in\Theta}b^m_{\mathcal{P}(n)}(\theta)\,
\rho_\theta\big(\mathbf{u}^{o_m}_n\mid\mathbf{x}^{o_m}_{\mathcal{P}(n)},\mathbf{x}^e_{\mathcal{P}(n)}\big), \\[2pt]
\omega_n&=\omega_{\mathcal{P}(n)}\,
\frac{\bar\rho_n}{\sum_{n'\in\mathcal{C}(\mathcal{P}(n))}\bar\rho_{n'}},
\end{aligned}
\label{eq:node_weight}
\end{equation}
for $n\!\in\!\mathbb{N}\setminus\{n_0\}$, so that $\omega_n$ is the probability of reaching node $n$ and the weights sum to one at every depth of the tree, in particular over the leaves, $\sum_{n\in\mathbb{L}}\omega_n\!=\!1$. Hence, each node enters the cost with the probability of the sampled opponent reactions that lead to it: a plausible reaction weighs heavily in the optimization, whereas an improbable one weighs little.

Collecting the terms above, the control objective over the scenario tree is
\begin{equation}
\begin{aligned}
J=&\sum_{n\in\mathbb{N}\setminus\mathbb{L}} \omega_n\,J_n(\mathbf{e}_n,\mathbf{u}^e_n)
+\sum_{n\in\mathbb{L}} \omega_n\,J^F_n(\mathbf{e}_n) \\
&+\sum_{n\in\mathbb{N}\setminus\{n_0\}} \omega_n\big(J^{\mathrm{pref}}_n(\ell^e_n)
+J^{\mathrm{slack}}_n(\boldsymbol{\xi}_n)\big).
\end{aligned}
\label{eq:objective}
\end{equation}


\subsection{Complete Optimization Problem}
\label{subsec:full}
Collecting the objective and the constraints above, the mixed-integer dual MPC problem solved at every time step $t$ is the following mixed-integer nonlinear program (MINLP):
\begin{equation}
\begin{aligned}
& \min_{\mathbf{u}^e,\boldsymbol{\sigma}^e,\boldsymbol{\xi}}\; \; J\ \ \text{by}\ \eqref{eq:objective} \\[2pt]
\text{s.t.}\;\;  & \mathbf{x}^e_{n_0}=\mathbf{x}^e_t,\ \mathbf{x}^{o_m}_{n_0}=\mathbf{x}^{o_m}_t,\  \ell^e_{n_0}=\ell^e_t,\ b^m_{n_0}=b^m_t, \\
& \mathbf{x}^e_n \!=\! A\mathbf{x}^e_{\mathcal{P}(n)} \!+\! B\mathbf{u}^e_{\mathcal{P}(n)}, \; \mathbf{x}^{o_m}_n \!=\! A\mathbf{x}^{o_m}_{\mathcal{P}(n)} \!+\! B\mathbf{u}^{o_m}_n, \\
& \mathbf{u}^{o_m}_n=g\big(\mathbf{x}^{o_m}_{\mathcal{P}(n)},\mathbf{x}^e_{\mathcal{P}(n)};\theta^m_n\big)+\boldsymbol{\epsilon}^{o_m}_n, \\
& b^m_n=\Psi\big(b^m_{\mathcal{P}(n)},\mathbf{u}^{o_m}_n,\mathbf{x}^{o_m}_{\mathcal{P}(n)},\mathbf{x}^e_{\mathcal{P}(n)}\big), \\
& \ell^e_n=\ell^e_{\mathcal{P}(n)} +(\delta^{+}_n-\delta^{-}_n)\in\mathcal{L}, \quad \omega_n\ \text{by}\ \eqref{eq:node_weight}, \\
& \delta^{+}_n+\delta^{-}_n \le 1, \quad \textstyle\sum_{q\in\mathcal{Q}}\gamma^{q,m}_n=1, \\
& \underline{\eta}^{e}_t-\xi^{\mathrm{rd}}_n \;\le\; \eta^{e}_n \;\le\; \overline{\eta}^{e}_t+\xi^{\mathrm{rd}}_n, \\
& \mathbf a_q^{\top}\mathbf r^m_n \;\ge\; d_q - M\big(1-\gamma^{q,m}_n\big)-\xi^{\mathrm{sf},m}_n, \\
& \mathbf{x}^e_n\in\mathcal{X},\; \mathbf{u}^e_n\in\mathcal{U},\; \boldsymbol{\sigma}^e_n\in\{0,1\}^{2+4N_o},\; \boldsymbol{\xi}_n\ge0,
\end{aligned}
\label{eq:full_problem}
\end{equation}
where the recursions of the states, opponent inputs, beliefs, and lane index hold at $n\!\in\!\mathbb{N}\setminus\{n_0\}$, and the remaining constraints at $n\!\in\!\mathbb{N}$. Constraints indexed by $m$ or $q$ hold for every opponent $m\!\in\!\{1,\dots,N_o\}$ and region $q\!\in\!\mathcal{Q}$. The tracking error $\mathbf{e}_n$ and the path-aligned quantities $\eta^{e}_n$, $v^{e}_{\tau,n}$, $\mathbf r^{m}_n$ are given by~\eqref{eq:track_error}, \eqref{eq:lateral_proj}, and~\eqref{eq:frame_offsets}, the road bounds $\underline{\eta}^{e}_t,\overline{\eta}^{e}_t$ by~\eqref{eq:road_bound}, and $\mathcal{X}$, $\mathcal{U}$ are the admissible state and input sets. The variables $\mathbf{u}^e$, $\boldsymbol{\sigma}^e$, and $\boldsymbol{\xi}$ collect $\mathbf{u}^e_n$, $\boldsymbol{\sigma}^e_n$, and $\boldsymbol{\xi}_n$ over the tree, whereas the states, lane indices, beliefs, and weights follow from them through the recursions. The sampled hidden parameters $\theta^m_n$ and noise realizations $\boldsymbol{\epsilon}^{o_m}_n$ of Sec.~\ref{subsec:tree}, together with the path-aligned frames $(\hat{\boldsymbol{\tau}}^{i}_k,\hat{\boldsymbol{\nu}}^{i}_k)$ and anchors $\mathcal{T}^{i}_k$, enter as constants fixed before the solve, so every path-aligned quantity is affine in the states, exactly on a straight path and to first order on a curved one.

This formulation can exhibit the \emph{dual-control effect}~\cite{mesbah2018stochastic}. The belief propagation along the tree captures whether an opponent's action is informative about the hidden parameters. The node weights~\eqref{eq:node_weight}, computed from the propagated beliefs, concentrate the objective on the plausible branches. Consequently, the optimization encourages ego behavior that sharpens the beliefs and fully exploits the belief state in planning. When the long-term gain outweighs the cost of probing, the ego elicits informative reactions from the opponents, which amounts to active information seeking.


\section{GNN-Guided Branch and Bound}

\begin{figure*}
    \centering
    \includegraphics[width=\textwidth]{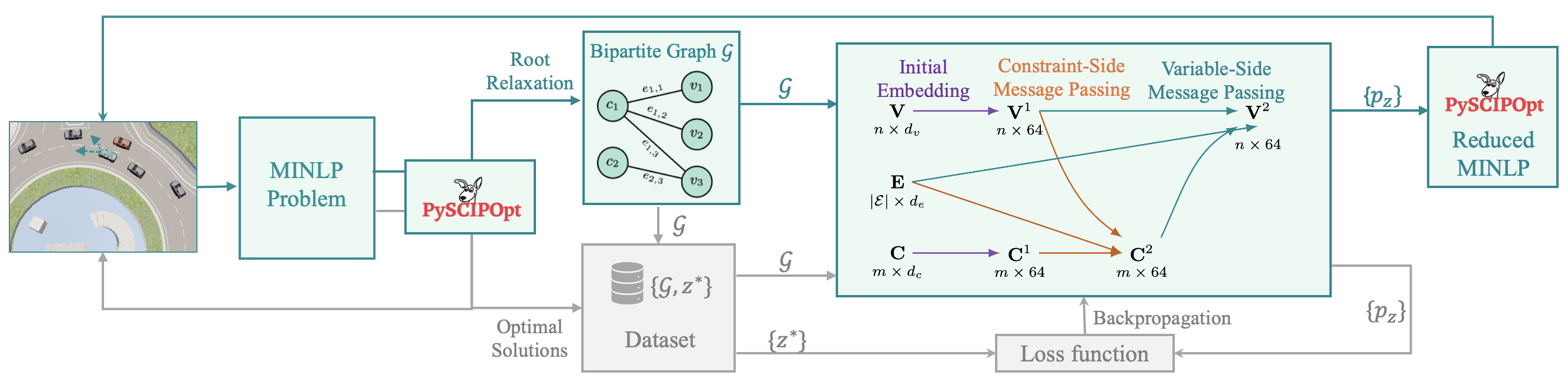}
    \caption{Solving the MINLP~\eqref{eq:full_problem} under GNN guidance. The online execution pipeline is shown in teal and the offline training in gray.}
     \label{fig:pipeline}
\end{figure*}

For the computationally demanding MINLP~\eqref{eq:full_problem}, standard solvers such as SCIP~\cite{SCIPOptSuite10}, which we use throughout, start from a root relaxation and close the optimality gap by B\&B, whose search cost grows quickly with the problem size. We therefore propose to use a GNN to predict the integer decisions and fix those predicted with high confidence, which reduces the search space left to the solver. The overall framework is shown in Fig.~\ref{fig:pipeline}.

\subsection{GNN Design and Training}
\subsubsection{Input Representation}
Following~\cite{gasse2019exact,khalil2022mip,nair2020solving}, we represent the MINLP with a bipartite graph $\mathcal G\!=\!(\mathbf{V} ,\mathbf{E}, \mathbf{C})$. The node sets $\mathbf{V}$ and $\mathbf{C}$ encode the properties of the variables and constraints, respectively, while the edge set $\mathbf{E}$ encodes the relationships between them. Therefore, $\mathcal G$ captures the structure of the optimization problem. This graph is extracted from the root relaxation of the MINLP, in which the integer decisions are treated as continuous.

\subsubsection{Output Representation} The GNN is designed to process the graph $\mathcal G$ and predict the optimal values of the binary decisions $\boldsymbol{\sigma}^e_n$ in~\eqref{eq:full_problem}. These binaries are constrained: at each node $n$, by~\eqref{eq:lane_recursion} at most one of the lane-change binaries $\delta_n^+$ and $\delta_n^-$ is active, and by~\eqref{eq:bigM} exactly one selector $\gamma_n^{q,m}$ is active for each opponent $m$. Rather than predicting the binaries one by one, we predict the underlying \emph{maneuver decisions}: for each node $n$, the signed lane change $\delta_n^+-\delta_n^-\in\{-1,0,+1\}$, and for each node $n$ and opponent $m$, the safe region $q\in\mathcal Q$ with $\gamma_n^{q,m}=1$. Let $\mathcal Z$ denote the set of maneuver decisions in problem~\eqref{eq:full_problem}, $z \!\in\! \mathcal Z$ one such decision, and $z^\star$ its value in an optimal solution of~\eqref{eq:full_problem}. For each decision $z$, the GNN predicts the distribution of its optimal value,
\begin{equation}
p_z(\zeta)=\Pr\big(z^\star=\zeta\big),
\label{eq:category_probability}
\end{equation}
where $\zeta$ ranges over the values $z$ can take, $\{-1,0,+1\}$ for a lane change and $\mathcal Q$ for a safe region. 


\subsubsection{GNN Architecture}
Following~\cite{gasse2019exact}, the GNN is designed as shown in Fig.~\ref{fig:pipeline}: given the input graph $\mathcal G$, the variable nodes $\mathbf{V}$ and the constraint nodes $\mathbf{C}$ are first embedded by separate multilayer perceptrons into $\mathbf{V}^{1}$ and $\mathbf{C}^{1}$, respectively. Next, in the constraint-side message passing, a multilayer perceptron computes a message for each edge from the embeddings of its two end nodes in $\mathbf{V}^{1}$ and $\mathbf{C}^{1}$ and its features in $\mathbf{E}$, and each constraint node aggregates the messages on its edges to update its embedding, yielding $\mathbf{C}^{2}$. An analogous variable-side pass then aggregates messages from the updated constraint nodes to obtain $\mathbf{V}^{2}$. Finally, a multilayer perceptron maps each variable embedding in $\mathbf{V}^{2}$ to a scalar score, and the distribution $p_z$ of a decision $z$ is the softmax over the scores of the binaries that compose it, with a zero score for the no-change option of a lane decision.

\subsubsection{Training Procedure} 
As shown by the gray offline pipeline in Fig.~\ref{fig:pipeline}, training data are collected from closed-loop simulations with randomized traffic; each data point consists of the graph $\mathcal G$ of one MINLP~\eqref{eq:full_problem} encountered in these simulations and the optimal values $\{z^\star\}_{z\in\mathcal Z}$ of its maneuver decisions, obtained by solving the full problem with SCIP~\cite{SCIPOptSuite10}. The GNN is trained by minimizing the cross-entropy $-\sum_{z\in\mathcal Z}\log p_z(z^\star)$ of the predicted distributions at these labels, summed over all training problems.


\subsection{GNN-Guided Variable Fixing}\label{subsec:fixing}

The distributions $\{p_z\}_{z\in\mathcal Z}$ predicted by the GNN are used to fix integer decisions, which results in a reduced optimization problem. For each decision $z$, the predicted value is $\widehat{z}=\operatorname*{arg\,max}_{\zeta}p_z(\zeta)$. If its probability satisfies $p_z(\widehat z)\geq\bar p$, the prediction is deemed confident enough to fix $z$ to $\widehat z$, and the corresponding entries of $\boldsymbol{\sigma}^e_n$ are fixed accordingly. Otherwise, the decision remains free and is left to the reduced MINLP. The threshold $\bar p$ sets the balance: raising it fixes fewer decisions and leaves more of them free, while lowering it reduces the search space further at a higher risk of fixing a decision wrongly. The reduced MINLP, obtained by substituting the fixed assignments into~\eqref{eq:full_problem}, is solved with the SCIP solver.

\section{Simulation Results and Validation}

The implementation details, source code, and video demonstrations are available on the project website\footnote{\hypersetup{urlcolor=black}\urlstyle{same}\url{https://anonymous.4open.science/api/repo/mixed-integer-dual-mpc-project-3CA9/file/index.html?v=3f7e589b}}. In the following case studies, the opponents follow the behavior model in~\eqref{eq:opp_mean} and~\eqref{eq:opponent_policy}, with $\theta\!\in\!\{\theta_{\mathrm{cau}},\theta_{\mathrm{agg}}\}$ per opponent.

\begin{figure}[ht]
 \centering
 \includegraphics[width=0.95\linewidth]{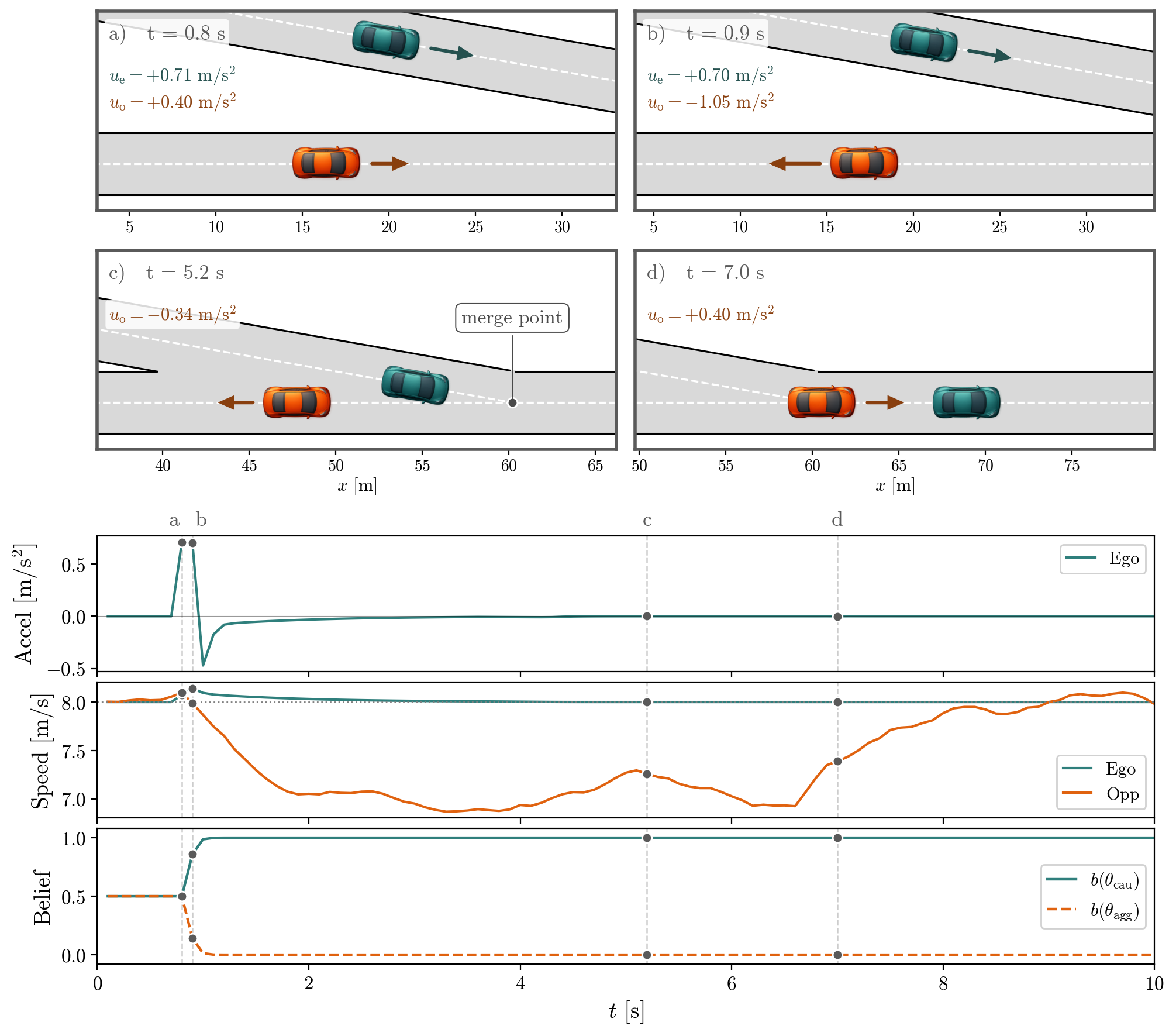}
 \caption{Representative closed-loop ramp-merging episode.
The marked snapshots show the evolution of the interaction, while the lower panels report the ego's emergent probing action, vehicle speeds, and belief convergence after probing. }
 \label{fig:ramp_result}
\end{figure}

\subsection{Case Study 1: Single-Opponent Ramp Merging}

To clearly illustrate the active probing behavior of the dual MPC controller, we first test it in a single-opponent, single-lane ramp-merging scenario. The ego vehicle (green) travels along the on-ramp and attempts to merge into the mainline, where a single opponent (orange) with an unknown hidden parameter $\theta\!\in\!\{\theta_{\mathrm{cau}},\theta_{\mathrm{agg}}\}$ approaches the merging point at roughly the same time as the ego. The ego decides whether to merge ahead of or behind the opponent, depending on the opponent's latent willingness to yield.

Fig.~\ref{fig:ramp_result} shows an episode that exhibits active probing. The MPC controller uses $H=15$, $H_b=2$, and $\Delta t=0.1\,\mathrm{s}$. Before the interaction, the ego does not know the opponent's true hidden parameter, and the belief remains neutral. At $t=0.8\,\mathrm{s}$, driven by the dual-control effect, the ego accelerates to elicit an informative response from the opponent. This maneuver sacrifices short-term tracking performance but is expected to reduce the long-term cost by reducing the uncertainty about $\theta$. Once the ego observes that the opponent yields, its belief converges to the cautious mode. Consequently, at $t=5.2\,\mathrm{s}$, the ego holds the desired speed $v^{e}_{\mathrm{des}}$ without braking and confidently merges ahead. 
This episode illustrates how proactive information seeking reduces uncertainty about the opponent’s intention and supports a more informed merging decision.

\begin{figure*}[htb]
    \centering
    \includegraphics[width=\linewidth,trim={10pt 0 0 0},clip]{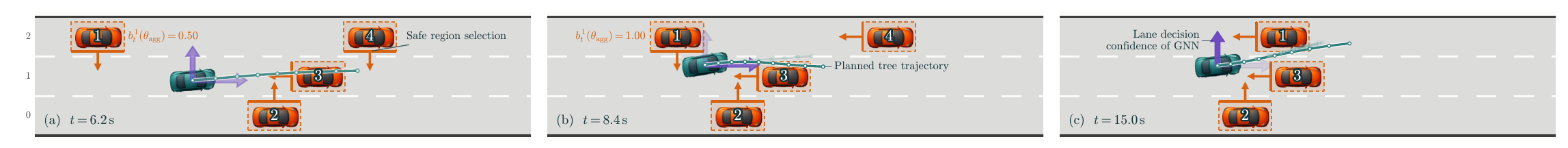}
    \caption{Three snapshots of the highway case study. Orange arrows mark the safe-region selections, the purple arrows mark the candidate lane decisions, each shaded by the GNN confidence, and the teal polyline traces the planned trajectory along the scenario tree.}
    \label{fig:highway} 
\end{figure*}

\subsection{Case Study 2: Multi-Lane Multi-Opponent Highway}


As shown in Fig.~\ref{fig:highway}, this case study considers a three-lane highway with $8$--$20$ surrounding vehicles traveling at $6$--$8\,\mathrm{m/s}$. Their initial states and hidden parameters $\theta$ are sampled at random. To induce overtaking interactions, the ego vehicle's desired speed is set to $v^{e}_{\mathrm{des}}=10\,\mathrm{m/s}$, above that of the traffic. At each MPC step, up to $5$ surrounding vehicles are included as opponents in the optimization. With $H=8$, $\Delta t=0.2\,\mathrm{s}$, and $H_b=2$, the scenario tree contains $31$ nodes and $819$ integer variables. The GNN is trained on $1000$ MINLP instances collected in the highway scenario and tested on $411$ instances from newly generated scenarios. The confidence threshold is set to $\bar p=0.95$. With this threshold, the fixing accuracy on the test set is $99.95\%$, that is, only $0.05\%$ of the fixed decisions differ from their optimal values. All computations are run on an Apple M1 Pro processor. 



\subsubsection{Interpretation of the Fixed Decisions}
Fig.~\ref{fig:highway} illustrates how GNN guidance reduces the number of free integer variables considered by the optimizer. The purple and orange arrows mark the lane-change and safe-region decisions predicted by the GNN at the current step, respectively. In panel~(a), the GNN leans toward a change to lane~2, but with low confidence, so the decision is left to the B\&B search, and the optimized trajectory probes slightly toward lane~2. As $V^{o_1}$ accelerates to close the gap instead of yielding, the belief $b^{1}_t(\theta_{\mathrm{agg}})$ converges to one, and in panel~(b) the GNN prediction shifts toward staying in lane~1. Once $V^{o_1}$ has passed, in panel~(c), the GNN again predicts the change to lane~2, and the ego carries it out. 
In these three snapshots, the lane-change decision is never fixed by the GNN, whereas the safe-region selections are fixed in all three snapshots, which saves solve time. The GNN thus indicates which decisions are critical in a given situation and which are not. This interpretation reflects an understanding of the interactive scenario, which is valuable for safety-critical applications. Note that the snapshots show only the current-step decisions, while the GNN predictions are less confident at deeper nodes. Over all instances, $76.3\%$ of the integer decisions are fixed at $\bar p=0.95$.

\subsubsection{Speedup and Solution Quality}
\begin{figure}[htb]
    \centering
    \includegraphics[width=0.85\linewidth]{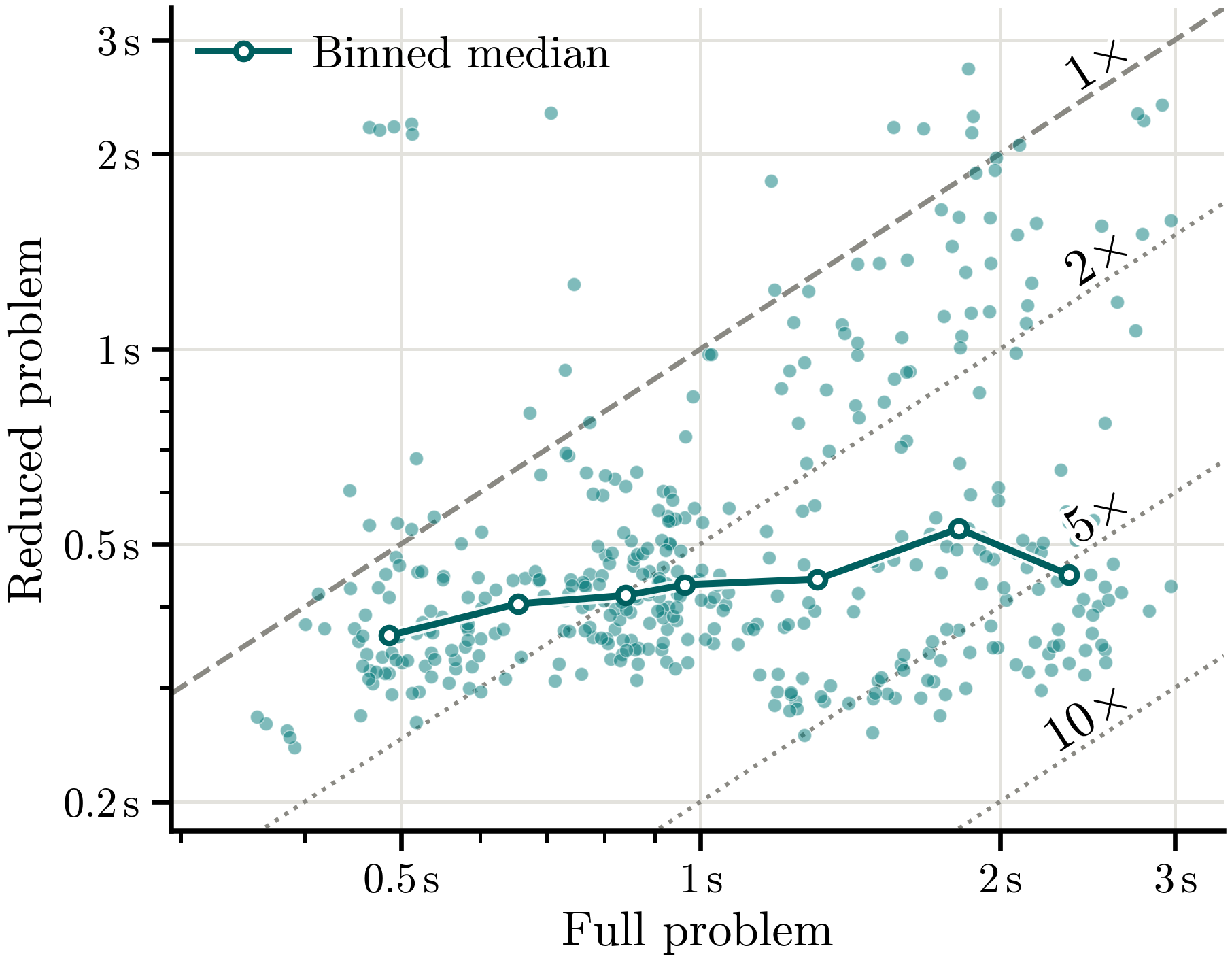}\caption{Solve time of the GNN-reduced problem against that of the full problem, on logarithmic axes. The dashed diagonal marks equal solve time ($1\times$) and the dotted diagonals $2\times$, $5\times$, and $10\times$ speedups; the teal curve is the median within equal-count bins.} 
    \label{fig:speedup}
\end{figure}

To evaluate the acceleration by GNN-guided fixing, Fig.~\ref{fig:speedup} plots the solve time of the reduced problem after fixing against that of the full problem~\eqref{eq:full_problem} for the MINLP instances collected from $13$ trials of the highway case study. In $94.4\%$ of these instances, the reduced solve finishes faster than the full solve. The full solve times range from approximately $0.4$ to $3\,\mathrm{s}$ across the traffic situations, whereas the reduced solve times concentrate around $0.3$--$0.65\,\mathrm{s}$. GNN-guided fixing reduces the solve time by $2.5\times$ on average and by up to $10\times$ on the most demanding instances. The binned median curve stays nearly flat as the full solve time grows, which indicates that GNN-guided fixing mitigates the additional computational burden of the harder instances. Since GNN inference itself takes only a few milliseconds, this GNN guidance leads to a substantial saving in overall solve time, which is crucial for real-time deployment. Moreover, the speedup does not come at the expense of solution quality: on the test set, GNN fixing reaches a fixing accuracy of $99.95\%$. Over the $411$ test instances, GNN fixing degrades the objective by at most $0.0080\%$, so the reduced solve stays near-optimal. Despite the significant speedup, the reduced solve times still exceed $\Delta t=0.2\,\mathrm{s}$ on the Apple M1 Pro, so the controller does not yet run in real time. Hardware acceleration and parallel processing are left for future work.

\subsection{Case Study 3: Multi-Lane Multi-Opponent Roundabout}
Finally, we evaluate the proposed controller in a multi-lane roundabout in the high-fidelity simulator CARLA~\cite{dosovitskiy2017carla}. Unlike the previous case studies on straight roads, where the path-aligned frame $(\hat{\boldsymbol{\tau}}^{i}_k,\hat{\boldsymbol{\nu}}^{i}_k)$ is constant along the horizon, the frame rotates around the roundabout with the arc length, so the path-aligned frames and lane-reference geometry are adapted to the curved road. As shown in Fig.~\ref{fig:roundabout}, the ego vehicle (green) enters the roundabout from the bottom and aims for the exit on the upper left. To track the desired speed $v^{e}_{\mathrm{des}}$, the ego first changes to the inner lane to avoid the slower vehicle ahead. It then encounters a circulating opponent (orange) with an unknown hidden parameter $\theta\!\in\!\{\theta_{\mathrm{cau}},\theta_{\mathrm{agg}}\}$, which may or may not yield to the ego. Driven by the dual-control effect, the ego actively probes the opponent to elicit an informative response. When the opponent is cautious, it yields, the belief converges to the cautious value, and the ego confidently exits the roundabout. When the opponent is aggressive, it does not yield; the ego therefore defers the exit and completes another lap before leaving the roundabout. The ego thus handles the interaction on the multi-lane, multi-opponent roundabout proactively, demonstrating the benefit of integer decisions in the dual MPC framework.

\section{Conclusion}
To handle realistic traffic, in which the reactions of multiple opponents on multi-lane roads exhibit multi-modal uncertainty driven by hidden parameters, this paper presented a mixed-integer dual MPC framework for interactive driving. The formulation incorporates integer variables to handle the combinatorial nature of the interaction and generates proactive behavior by accounting for both tracking performance and uncertainty reduction. To address the increased computational cost of the resulting mixed-integer program, an interpretable GNN-guided variable-fixing scheme was proposed that fixes the high-confidence maneuver decisions and leaves only the ambiguous ones to the solver. This scheme reduces the solve time substantially and flattens its growth with the complexity of the interaction.

\bibliographystyle{IEEEtran} 
\bibliography{ref} 

\end{document}

%% file: drawing/example.tex
\begin{tikzpicture}[
    font=\small,
    >=Latex,
    tnode/.style={circle, draw=black, thick, fill=nodecol, inner sep=0pt,
                  minimum size=6.5mm, font=\footnotesize},
    sedge/.style={-Stealth, thick, black},
    lab/.style={font=\small, text=black},
    plab/.style={above=1pt, fill=white, inner sep=1pt}
]
\node[tnode] (n0) at (0,0) {$n_0$};
\node[tnode] (n1) at (2.0, 0.85) {$n_1$};
\node[tnode] (n2) at (2.0,-0.85) {$n_2$};
\node[tnode] (n3) at (4.0, 1.4) {$n_3$};
\node[tnode] (n4) at (4.0, 0.45) {$n_4$};
\node[tnode] (n5) at (4.0,-0.45) {$n_5$};
\node[tnode] (n6) at (4.0,-1.4) {$n_6$};
\node[tnode] (n7)  at (6.0, 1.4) {$n_7$};
\node[tnode] (n8)  at (6.0, 0.45) {$n_8$};
\node[tnode] (n9)  at (6.0,-0.45) {$n_9$};
\node[tnode] (n10) at (6.0,-1.4) {$n_{10}$};
\node[tnode] (n11) at (8.0, 1.4) {$n_{11}$};
\node[tnode] (n12) at (8.0, 0.45) {$n_{12}$};
\node[tnode] (n13) at (8.0,-0.45) {$n_{13}$};
\node[tnode] (n14) at (8.0,-1.4) {$n_{14}$};
\draw[green!55!black, dashed, thick] (8.0,0) ellipse (0.6 and 1.9);
\node[lab, text=green!55!black, anchor=west] at (8.68,0) {$\mathbb{L}$};
\draw[sedge] (n0) -- node[lab,above,sloped] {$\mathbf{u}^{o}_{n_1}$} (n1);
\draw[sedge] (n0) -- node[lab,below,sloped] {$\mathbf{u}^{o}_{n_2}$} (n2);
\draw[sedge] (n1) -- node[lab,above,sloped] {$\mathbf{u}^{o}_{n_3}$} (n3);
\draw[sedge] (n1) -- node[lab,below,sloped] {$\mathbf{u}^{o}_{n_4}$} (n4);
\draw[sedge] (n2) -- node[lab,above,sloped] {$\mathbf{u}^{o}_{n_5}$} (n5);
\draw[sedge] (n2) -- node[lab,below,sloped] {$\mathbf{u}^{o}_{n_6}$} (n6);
\draw[sedge] (n3) -- node[lab,plab,pos=0.62] {$\mathbf{u}^{o}_{n_7}$} (n7);
\draw[sedge] (n4) -- node[lab,plab,pos=0.62] {$\mathbf{u}^{o}_{n_8}$} (n8);
\draw[sedge] (n5) -- node[lab,plab,pos=0.62] {$\mathbf{u}^{o}_{n_9}$} (n9);
\draw[sedge] (n6) -- node[lab,plab,pos=0.62] {$\mathbf{u}^{o}_{n_{10}}$} (n10);
\draw[sedge] (n7)  -- node[lab,plab] {$\mathbf{u}^{o}_{n_{11}}$} (n11);
\draw[sedge] (n8)  -- node[lab,plab] {$\mathbf{u}^{o}_{n_{12}}$} (n12);
\draw[sedge] (n9)  -- node[lab,plab] {$\mathbf{u}^{o}_{n_{13}}$} (n13);
\draw[sedge] (n10) -- node[lab,plab] {$\mathbf{u}^{o}_{n_{14}}$} (n14);
\draw[decorate, decoration={brace, amplitude=5pt}, thick]
    (-0.5,0.98) -- (-0.5,-0.98);
\node[lab, anchor=east, align=right] at (-0.66,0)
    {$\mathbf{x}^e_{n_0}\!=\!\mathbf{x}^e_t$\\
     $\mathbf{x}^{o_m}_{n_0}\!=\!\mathbf{x}^{o_m}_t$\\[1pt]
     $b^m_{n_0}\!=\!b^m_t$\\[1pt]
     $\ell^e_{n_0}\!=\!\ell^e_t$\\[1pt]
     \textcolor{egocol}{$\mathbf{u}^e_{n_0},\ \boldsymbol{\sigma}^e_{n_0}$}};
\draw[dashed, gray] (5.0,-2.0) -- (5.0,2.3);
\draw[decorate, decoration={brace, amplitude=4pt}, thick]
    (0,2.4) -- (4.9,2.4)
    node[midway, above=4pt, lab, align=center] {branching horizon\\($H_b\!=\!2$)};
\draw[decorate, decoration={brace, amplitude=4pt}, thick]
    (5.1,2.4) -- (8.0,2.4)
    node[midway, above=4pt, lab, align=center] {propagation horizon\\($H\!-\!H_b\!=\!2$)};
\foreach \k/\x in {0/0, 1/2.0, 2/4.0, 3/6.0, 4/8.0}
    {\node[lab] at (\x,-2.25) {$k\!=\!\k$};}
\end{tikzpicture}